%% file: main_VLDB.tex
\documentclass[sigconf, nonacm]{acmart}

\newcommand\vldbdoi{XX.XX/XXX.XX}
\newcommand\vldbpages{XXX-XXX}
\newcommand\vldbvolume{14}
\newcommand\vldbissue{1}
\newcommand\vldbyear{2020}
\newcommand\vldbauthors{\authors}
\newcommand\vldbtitle{\shorttitle} 
\newcommand\vldbavailabilityurl{URL_TO_YOUR_ARTIFACTS}
\newcommand\vldbpagestyle{plain} 

\usepackage{mathptmx}  
\usepackage[T1]{fontenc}
\usepackage[utf8]{inputenc}
\usepackage{pslatex}
\usepackage{breakurl}           
\usepackage{makecell}
\usepackage{url}                
\usepackage{xcolor}             
\usepackage[]{hyperref}         
\hypersetup{                    
  colorlinks,
  linkcolor={green!80!black},
  citecolor={red!70!black},
  urlcolor={blue!70!black}
}
\usepackage [ruled,linesnumbered] {algorithm2e}
\usepackage[noend]{algorithmic}
\definecolor{green}{rgb}{0,0.6,0}
\definecolor{gray}{rgb}{0.5,0.5,0.5}
\definecolor{mauve}{rgb}{0.58,0,0.82}

\usepackage{amsmath}
\usepackage{pgfplots}
\pgfplotsset{compat=newest}
\usepackage{filecontents}

\usepackage{caption}
\usepackage{subcaption}               
\usepackage{tabularx}                 
\usepackage{colortbl}                 
\usepackage{multirow}                 
\usepackage{listings}                 
\usepackage{tikz}                     
\usepackage{graphicx}                 
\graphicspath{ {images/} }            
\usepackage{todonotes}                
\usepackage{hyperref}                 
\usepackage{url}
\usepackage{pgfplots, pgfplotstable}
\pgfplotsset{compat=1.8}
\usepgfplotslibrary{statistics}

\usepackage{tikz}
\usepackage{tikz, tkz-graph}
\usetikzlibrary{chains}
\usetikzlibrary{arrows}
\usetikzlibrary{positioning}
\usetikzlibrary{shapes.geometric}
\usetikzlibrary{calc}
\usetikzlibrary{chains}
\usetikzlibrary{arrows}
\usetikzlibrary{positioning}
\usetikzlibrary{shapes.geometric}
\usetikzlibrary{calc}
\usetikzlibrary{automata}
\usetikzlibrary{scopes}
\usetikzlibrary{backgrounds}
\usetikzlibrary{patterns}
\usetikzlibrary{decorations.pathreplacing}
\usetikzlibrary{arrows, decorations.markings}
\tikzset{
	-|-/.style={
		to path={
			(\tikztostart) -| ($(\tikztostart)!#1!(\tikztotarget)$) |- (\tikztotarget)
			\tikztonodes
		}
	},
	-|-/.default=0.5,
	|-|/.style={
		to path={
			(\tikztostart) |- ($(\tikztostart)!#1!(\tikztotarget)$) -| (\tikztotarget)
			\tikztonodes
		}
	},
	|-|/.default=0.5,
	-|-|/.style 2 args={
		to path={
			(\tikztostart) -| ($(\tikztostart)!#1!(\tikztotarget)$) |- ($(\tikztostart)!#2!(\tikztotarget)$) -| (\tikztotarget)
			\tikztonodes
		}
	},
	-|-|/.default=0.5,
}

\tikzstyle{vecArrow} = [thick, decoration={markings,mark=at position
	1 with {\arrow[semithick]{open triangle 60}}},
double distance=1.4pt, shorten >= 5.5pt,
preaction = {decorate},
postaction = {draw,line width=1.4pt, white,shorten >= 4.5pt}]
\tikzstyle{innerWhite} = [semithick, white,line width=1.4pt, shorten >= 4.5pt]

\tikzset{
	invisible/.style={opacity=0},
	visible on/.style={alt={#1{}{invisible}}},
	alt/.code args={<#1>#2#3}{%
		\alt<#1>{\pgfkeysalso{#2}}{\pgfkeysalso{#3}} 
	},
}

\makeatletter
\pgfplotsset{
    boxplot prepared from table/.code={
        \def\tikz@plot@handler{\pgfplotsplothandlerboxplotprepared}%
        \pgfplotsset{
            /pgfplots/boxplot prepared from table/.cd,
            #1,
        }
    },
    /pgfplots/boxplot prepared from table/.cd,
        table/.code={\pgfplotstablecopy{#1}\to\boxplot@datatable},
        row/.initial=0,
        make style readable from table/.style={
            #1/.code={
                \pgfplotstablegetelem{\pgfkeysvalueof{/pgfplots/boxplot prepared from table/row}}{##1}\of\boxplot@datatable
                \pgfplotsset{boxplot/#1/.expand once={\pgfplotsretval}}
            }
        },
        make style readable from table=lower whisker,
        make style readable from table=upper whisker,
        make style readable from table=lower quartile,
        make style readable from table=upper quartile,
        make style readable from table=median,
        make style readable from table=lower notch,
        make style readable from table=upper notch,
        make style readable from table=draw position,
        make style readable from table=draw,
}
\makeatother

\usepackage{bbding}
\usepackage{pifont}
\usepackage{wasysym}

\usepackage{cleveref}    
\crefformat{section}{\S#2#1#3} 
\crefformat{subsection}{\S#2#1#3}
\crefformat{subsubsection}{\S#2#1#3}

\graphicspath{{./figure/}}

\usepackage[title]{appendix}
\begin{document} 
\newcommand{\sysname}{FpgaHub} 

\date{}

\title{\sysname: Fpga-centric Hyper-heterogeneous Computing Platform for Big Data Analytics}

\author{Zeke Wang, Jie Zhang, Hongjing Huang, Yingtao Li, Xueying Zhu, Mo Sun, Zihan Yang, De Ma, Huajing Tang, Gang Pan, Fei Wu, Bingsheng He$^{\star}$, Gustavo Alonso$^{**}$}
\affiliation{%
  \institution{Zhejiang University, China $ $ $^{\star}$National University of Singapore, Singapore $ $ $^{**}$Systems Group, ETH Zurich, Switzerland}
}



\input{content/0-abstract.tex}

\maketitle

\pagestyle{\vldbpagestyle}
\begingroup\small\noindent\raggedright\textbf{PVLDB Reference Format:}\\
\vldbauthors. \vldbtitle. PVLDB, \vldbvolume(\vldbissue): \vldbpages, \vldbyear.\\
\href{https://doi.org/\vldbdoi}{doi:\vldbdoi}
\endgroup
\begingroup
\renewcommand\thefootnote{}\footnote{\noindent
This work is licensed under the Creative Commons BY-NC-ND 4.0 International License. Visit \url{https://creativecommons.org/licenses/by-nc-nd/4.0/} to view a copy of this license. For any use beyond those covered by this license, obtain permission by emailing \href{mailto:info@vldb.org}{info@vldb.org}. Copyright is held by the owner/author(s). Publication rights licensed to the VLDB Endowment. \\
\raggedright Proceedings of the VLDB Endowment, Vol. \vldbvolume, No. \vldbissue\ %
ISSN 2150-8097. \\
\href{https://doi.org/\vldbdoi}{doi:\vldbdoi} \\
}\addtocounter{footnote}{-1}\endgroup

\ifdefempty{\vldbavailabilityurl}{}{
\vspace{.3cm}
\begingroup\small\noindent\raggedright\textbf{PVLDB Artifact Availability:}\\
The source code, data, and/or other artifacts have been made available at \url{\vldbavailabilityurl}.
\endgroup
}

\vspace{-1ex}
\input{content/1-introduction.tex}

\vspace{-1ex}
\input{content/2-system.tex}

\vspace{-1ex}
\input{content/3-Design.tex}

\vspace{-1ex}
\input{content/4-experiment.tex}
\input{content/5-conclusion.tex}
\bibliographystyle{ACM-Reference-Format}
\bibliography{myref}

\end{document}

%% file: content/0-abstract.tex
\begin{abstract}

Modern data analytics requires a huge amount of computing power and processes a massive amount of data. At the same time, the underlying computing platform is becoming much more heterogeneous on both hardware and software.  
Even though specialized hardware, e.g., FPGA- or GPU- or TPU-based systems, often achieves better performance than a CPU-only system due to the slowing of Moore's law, such systems are limited in what they can do. For example, GPU-only approaches suffer from severe IO limitations. 
To truly exploit the potential of hardware heterogeneity, we present \sysname, an FPGA-centric hyper-heterogeneous computing platform for big data analytics. The key idea of \sysname{} is to use reconfigurable computing to implement a versatile hub complementing other processors (CPUs, GPUs, DPUs, programmable switches, computational storage, etc.). Using an FPGA as the basis, we can take advantage of its highly reconfigurable nature and rich IO interfaces such as PCIe, networking, and on-board memory, to place it at the center of the architecture and use it as a data and control plane for data movement, scheduling, pre-processing, etc. \sysname{} enables architectural flexibility to allow exploring the rich design space of heterogeneous computing platforms.  

%




\end{abstract}

%% file: content/1-introduction.tex
\section{Introduction}
\vspace{-1ex}
High-speed data generation has been a prominent trend in recent years in areas such as E-commerce~\cite{data_analytics_electronic16}, AI~\cite{waswani2017attention}, IoT~\cite{mohammadi2018deep}, etc. The data deluge offers both opportunities and challenges. The potential comes from the ability to gain real-time insights and improved decision-making from large data collections. 
The challenge arises from the massive computing power needed to process all the data. 
Due to the slowing down of Moore's Law, the performance of a single CPU has hardly increased in the last years. Thus, data analytics has resorted to either scale horizontally by using many CPUs \cite{Spark, Flink, Databricks, Snowflake}, or to scale vertically through heterogeneous devices~\cite{h100, u280, shen2016darwin, ma2024darwin3, aqua, boost}. With the advent of AI~\cite{liu2024deepseek, deepseekv3tr_arxiv24, achiam2023gpt, dubey2024llama}, the underlying computing platform is rapidly evolving towards a collection of highly specialized devices (GPUs, DPUs, TPUs, smart NICs, etc.). 
Often, however, many new systems in research and industry use a single accelerator in combination with a CPU and thus cannot fully provide a perfect match for a particular application. For example, NVIDIA produces and advocates GPU-only training and inference systems that have higher computing power, memory capacity, inter-GPU bandwidth, and CPU-GPU bandwidth. 
In the following, we analyze the advantages and disadvantages of leveraging a single heterogeneous device for large-scale analytics. 

\noindent{\bf GPU-only Data Analytics. } A broad range of existing work~\cite{join_gpu_sigmod08, mars_pact08, gpudb_vldb23} leverages GPUs to accelerate big data analytics. This offers several advantages and some disadvantages. 
The {\it advantages} of GPU-only data analysis are massive processing power and huge memory bandwidth that are badly needed by compute- and memory-intensive data analysis, such as deep learning tasks. For example, the newest NVIDIA H100 GPU provides 989 TFLOPS FP32 compute power and up to 3.35 TB/s memory bandwidth~\cite{h100}. 
Its main {\it disadvantage} lies in low communication ability. The leading collectives, i.e., NCCL~\cite{nccl_nvidia_18}, rely on a significant amount of GPU computing resources (e.g., 20 out of 132 SMs in the H80 GPU) and memory states to implement.\footnote{NVIDIA SHARP~\cite{sharp_ISC20} still needs GPU resources to cooperate, while NVIDIA does not suggest using its DPU, e.g., BlueField, to accelerate collectives. 20 out of 132 SMs are used for both IB and NVLink} 
However, the cost-efficient DeepSeek AI system suggests offloading the entire collectives to a network co-processor, i.e., SmartNIC, to 1) migrate the GPU resources needed by collectives and 2) fully overlap collectives and computation~\cite{deepseekv3tr_arxiv24}. 
We conclude that GPU-only collectives become the main limiting factor for GPUs in LLM training and inference. 

\noindent{\bf Programmable Switch-only Data Analytics. } Programmable switches, e.g., P4 switch~\cite{jin2018netchain, lerner2019netqueryprocessing, jin2017netcache}, have been used to accelerate applications in distributed settings. The main {\it advantage} is the ability to process data in-flight, e.g., aggregation and broadcast~\cite{p4_hotos19, switchml,atp_nsdi21}, which can highly reduce the end host's network processing requirements. 
To achieve this, each coordinated host CPU typically leverages DPDK or RDMA~\cite{switchml, pegasus_osdi20, eris_sosp17} to implement a custom host stack to communicate with the programmable switch. The main {\it disadvantage} is the high overhead from the CPU-initialized network stack which can easily become the performance bottleneck, because the compute capacity of the host CPU can typically not cope with increasing network bandwidth and higher network packet rates. 



\noindent{\bf FPGA-only Data Analytics. }A broad range of existing work uses FPGAs as a specialized accelerator for big data analytics, such as in database engines~\cite{histogram_sigmod14, Ibex_vldb14, Glacier_sigmod10, streams_vldb09, data_processing_vldb09, query_opencl_fpga_fpl16, partitioning_opencl_fpl15, pattern_matching_sigmod17}, and Machine Learning~\cite{ssimd_fpt23, mlweaving_tr, sparseACC_tcad24, columnml_vldb19, p4sgd, distRec_fpl21, vecsearch_sc23}. 
The {\it advantages} are two-fold. First, FPGAs enable the construction of custom dataflow for a particular application, such that the tailored application logic allows deep pipelining and does not suffer from the overhead associated with traditional general-purpose CPUs relying on costly instructions. %
Second, they can be used as a SmartNIC~\cite{alonso2019dpi, fpganic_atc22, strom_eurosys20, dpdpu_cidr25, smartds_isca23} to enable architectural flexibility to offload the application logic from the host CPU to the SmartNIC such that the application logic could directly interact with the network or storage without traversing the corresponding software layers. 
The main {\it disadvantages} are two-fold. First, it has low programmability and long synthesis time. Second, FPGAs do not have the highest computing power, memory capacity, or networking ability, compared with the best accelerator using the same MOSFET technology. Thus, using FPGAs as a pure accelerator does not necessarily yield the highest performance for big data analytics. 

In summary, these are just examples of the potential but also the challenges of using a single type of device beyond CPUs for big data analytics. In this paper, we argue that the true potential of heterogeneous architectures requires a different approach. In particular, the way FPGAs can be used has not been fully explored. To this end, we present \sysname, an FPGA-centric hyper-heterogeneous computing platform intended to serve as the control and data plane of a heterogeneous system to address the limitations of each particular device by complementing their functionality while providing additional support for tasks such as LLM training, data routing, data pre-processing, scheduling, etc. An FPGA is the right device for such a task given its reconfigurable nature and the wealth of interfaces it supports, enabling line rata data processing as data flows between devices and storage. 
Our initial experimental results validate the potential of \sysname{} to complement other heterogeneous devices. 



%% file: content/2-system.tex
\section{Potentials of \sysname}
\label{sec_potential}

\begin{figure}[t]
	\centering
	\includegraphics[width=7.5cm]{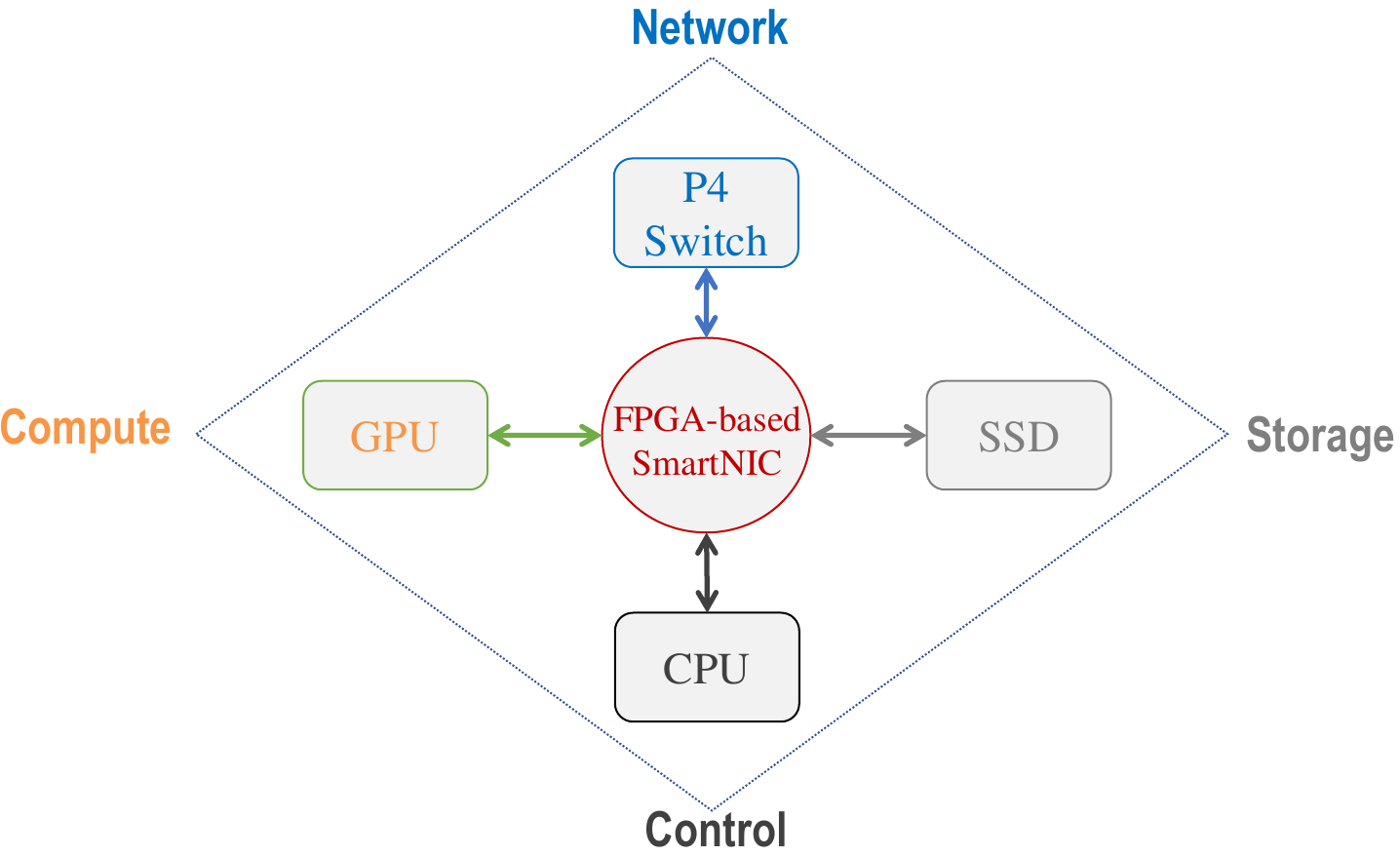}
	\vspace{-1ex}
	\caption{High-level overview of FPGA-centric hyper-heterogeneous computing platform}
	\vspace{-3ex}
	\label{fig_hyper_heterogeneous_platform}
\end{figure}

\vspace{-0.5ex}
In this section, we discuss the potential of our hyper-heterogeneous computing platform for big data analytics. Our platform is centralized on an FPGA-based SmartNIC, and we explore its ability of co-optimization with any other heterogeneous device to address their corresponding issues, as shown in Figure~\ref{fig_hyper_heterogeneous_platform}.  

\vspace{-1ex}
\subsection{Centralized Role: FPGA}
\label{subsec_fpga}
\vspace{-0.5ex}

Though an FPGA does not have the highest computing power, memory capacity, networking ability, and controlling ability, we first argue that FPGA's main advantage is to complement other heterogeneous devices, due to its highly reconfigurable nature and rich IO interfaces such as PCIe, networking, and on-board memory. We believe that FPGA is one of the best glue logics to complement other heterogeneous architectures. 


\noindent{\bf Computing Power. }FPGA has a massive amount of reconfigurable LUTs, RAMs, and DSPs for implementing a bit-level custom hardware engine for a particular application, yielding high performance and high energy efficiency. For example, the VPK180 FPGA board has 3.2M LUTs, 41MB RAMs, and 14k DSPs~\cite{vpk180}, where each DSP allows a 25x18 multiplication per cycle. An FPGA design typically achieves a frequency of 200MHz. 

\noindent{\bf Memory Capacity and Bandwidth. }A typical FPGA features a few DDR channels, or even HBM stacks, to host application states. For example, the U280 FPGA board has two DDR4 channels whose size is 32GB and two HBM stacks whose size is 8GB, where DDR4 provides up to 38.4GB/s memory bandwidth while HBM provides up to 460GB/s memory bandwidth~\cite{shuhai_fccm20, shuhai_tc22}. 

\noindent{\bf Internal IO. }A typical FPGA acceleration card features a PCIe integrated block to provide scalable and reliable serial interconnect to other PCIe devices, e.g., GPU and SSD, in the same server~\cite{pcie}. Due to its pure hardware implementation, an FPGA-based PCIe IP core enjoys stable low latency while keeping high throughput when talking to other PCIe devices. For example, the VPK180 FPGA board features a PCIe 5.0 X8 IP block.

\noindent{\bf External IO. }A typical FPGA acceleration card always features an integrated Ethernet subsystem to directly talk to other Ethernet devices via networking~\cite{cmac}. For example, the VPK180 FPGA board features several integrated 600Gb Ethernet subsystems. 


\vspace{-1ex}
\subsection{How to Complement a GPU?}
\label{subsec_gpu}
\vspace{-0.5ex}

\subsubsection{Advantages of GPU over FPGA}
Compared with FPGA, GPU has two obvious advantages. First, GPU has an order of magnitude higher computing power than FPGA at the same generation. For example, AMD Ultrascale FPGA has 50 TFLOPs computing power~\cite{vpk180} while GPU has 500 TFLOPs~\cite{h100}. The underlying reason is that GPU typically uses better manufacturing technology than FPGA at the same generation. GPU is a customized ASIC design whose frequency is higher (typically more than 1.7GHz), while FPGA design can typically achieve 200MHz frequency because FPGA is originally designed for reconfiguration, which comes at the cost of at least an order of magnitude larger chip space for each component such as adder. Second, GPUs have higher programmability than FPGAs. In particular, GPUs allow programmers to use the high-level CUDA (Compute Unified Device Architecture) to program such that it is relatively easy to implement a function on GPU, while FPGA mainly relies on low-level hardware description languages (HDL) and thus suffers from a steep learning curve and long compiling time. 

\subsubsection{Disadvantages of GPU over FPGA} Compared with FPGAs, GPUs have two limitations. 
First, a GPU has lower architectural flexibility than an FPGA. In particular, a GPU can only act as a compute-intensive PCIe accelerator of the host CPU, specially because GPU's runtimes such as CUDA runs on the CPU and it is needed to launch every GPU kernel. Therefore, GPU's architectural position is easily constrained by the CPU. In contrast, FPGAs can run independently with architectural freedom.\footnote{Nevertheless, the FPGA also can act as a PCIe accelerator of the host CPU.} 
Second, the GPU has poor network ability due to two factors. First, the GPU has a weak memory consistency model such that it is difficult to handle network states. Second, network functionality typically does not expose enough parallelism for GPU to fully leverage its massive computing power, because the GPU is not originally designed for networking. Figure~\ref{fig_fpganic} shows that collectives and GEMMs co-exist in the GPU and thus may suffer from severe interference issues~\cite{deepseekv3tr_arxiv24}, in terms of SM and memory bandwidth. 

\subsubsection{How to Complement GPU with FPGA} 
\begin{figure}[t]
	\centering
	\includegraphics[width=8.5cm]{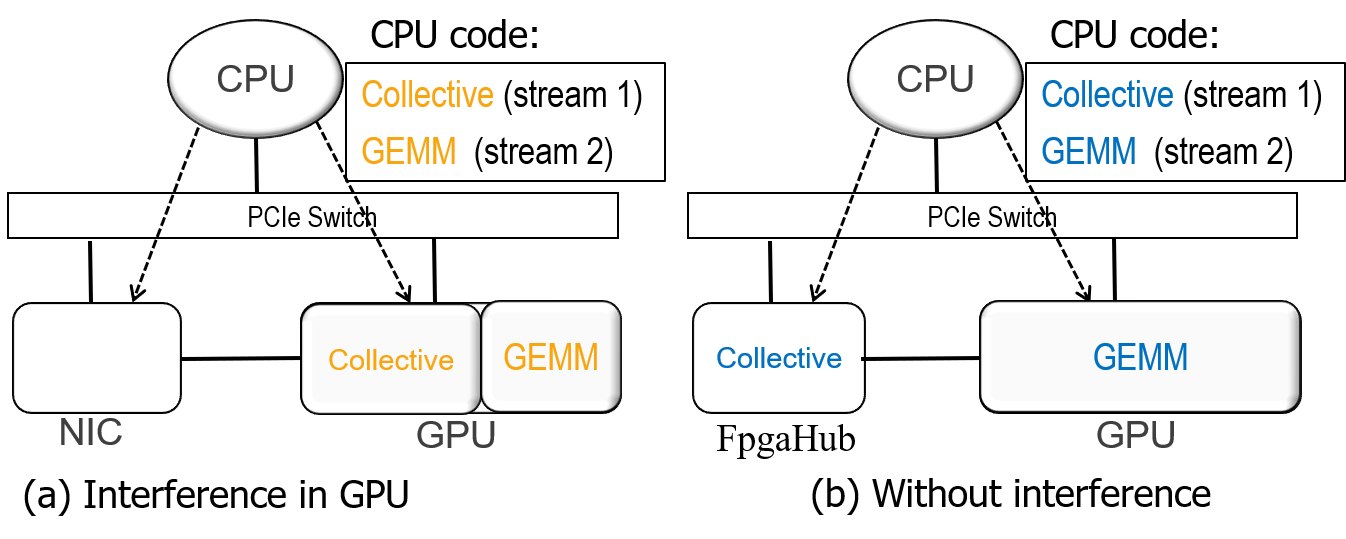}
	\vspace{-3ex}
	\caption{Comparison of with and w/o interference between collectives and GEMM}
	\vspace{-3ex}
	\label{fig_fpganic}
\end{figure}

Regarding GPU's limitation, we argue to set FPGA to be a GPU-centric SmartNIC, e.g., FpgaNIC, that enables control plane and data plane offloading~\cite{fpganic_atc22}, as shown in Figure~\ref{fig_fpganic}b. Data plane offloading enables FPGA to directly transfer data to GPU virtual memory via GPUDirect, without staging in CPU memory, while control plane offloading enables GPU to directly manipulate FPGA within a GPU kernel without CPU intervention. 
In the context of GPU-centric SmartNIC, we can offload networking functionality, e.g., collectives, from CPU/GPU to FPGA such that GPU only needs to focus on compute-intensive tasks, e.g., GEMM (General Matrix Multiplication), while FPGA focuses on network functionality, e.g., collective, to allow GPU to synchronous memory semantics to invoke. 
In particular, GPU can directly use one store instruction to trigger one doorbell register within the FPGA to start one collective operation. 
As such, \sysname{} enables GPU to have its own collective accelerator and thus enables an interesting interplay between network, GPU, and CPU. 

\subsection{How to Complement a P4 Switch?}
\label{subsec_p4}
\vspace{-0.5ex} 


Compared with FPGA, the main advantage of P4 Switch is that it flexibly processes packets with surprisingly high throughput. Specifically, Intel Wedge100-32x~\cite{barefoot-tofino} has 32 100Gbps network ports and supports 3.2Tbps packet processing at line rate. In contrast, FPGA (e.g., Xilinx U280~\cite{u280}) usually has single-digit (e.g., 2) 100Gbps network ports. 
%

\subsubsection{Three Limitations of P4 Switch}
Despite the advantages of P4 Switch, we identify that it has three limitations. 

First, it supports highly limited packet processing functions for two reasons. First, it has a small number (e.g., 12 for Wedge100-32x) of available pipe stages, which impedes the function with long data dependency paths from being implemented on the switch data plane~\cite{netlock}. Second, the switch data plane has limited small-scale ALUs, and thus can't support complex calculations like multiplication and division~\cite{p4survey, switch_hotnet24, p4_hotos19}.   

Second, it has limited on-chip storage resources. Specifically, a P4 switch only has tens of MBs of SRAM space to store massive states~\cite{sketchlib, rmt, silkroad, newton}. 

Third, we first identify that the traditional CPU-managed network transport cannot keep pace with a P4 switch~\cite{tiara}. 
Typically, each server needs a CPU-managed network transport to cooperate with the P4 switch and thus enter expensive CPU-centric coordination between servers, as shown in Figure~\ref{fig_transport}a. 
We identify that the P4 switch has significantly lower latency than the CPU-managed network transport. The intuition is that the pipeline latency of a packet that goes through all the stages of a P4 switch is roughly 1-2 us. However, a packet goes through a NIC to the CPU memory, where the lightweight CPU-managed network transport performs before sending a response back to the NIC. Such a round trip causes at least 10us latency.  


%

\subsubsection{How to Co-design P4 Switch with FPGA}
To address the above three limitations of P4 switch, we argue to complement P4 switch with FPGA-based SmartNIC to make the best of two worlds. 

First, we can leverage sufficient FPGA's programmable logics to implement massive complex calculations to consume line-rate network traffic. Our experience is that it costs less than 10\% resources of a modern datacenter FPGA to implement a compute-intensive kernel on 200Gbps network traffic~\cite{fpganic_atc22}. 

Second, we argue to offload states onto FPGA's on-board memory, because a typical FPGA features a few DDR channels, or even HBM stacks, to host massive application states.  

\begin{figure}[t]
	\centering
	\includegraphics[width=8.5cm]{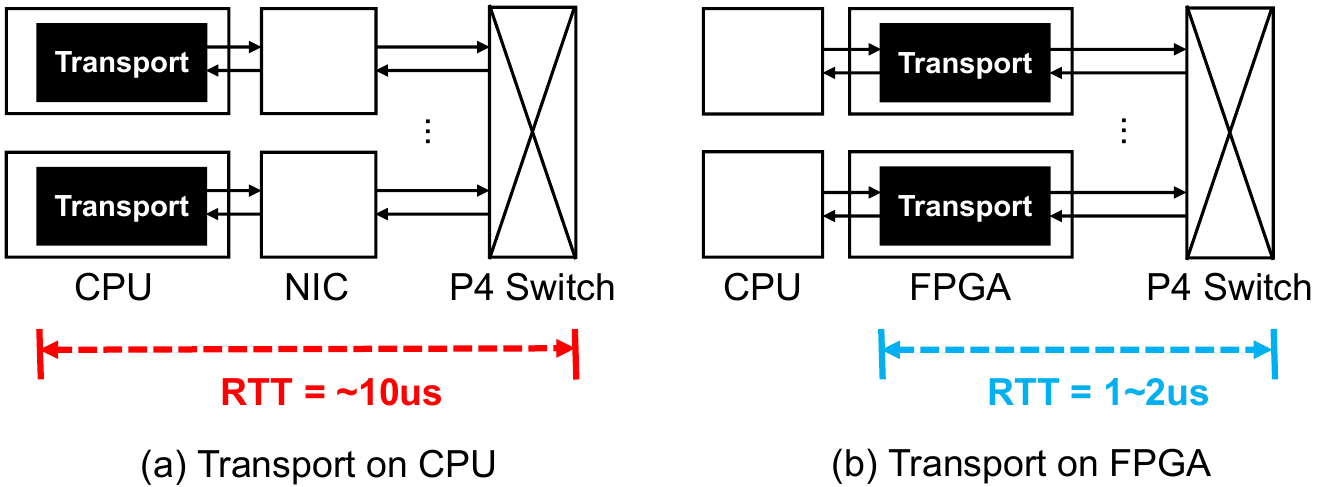}
	\vspace{-5ex}
	\caption{Comparison of CPU-based and FPGA-based transport.}
	\vspace{-3ex}
	\label{fig_transport}
\end{figure}

Third, we argue to offload network transports to FPGA, because FPGA can efficiently pipeline all the functionalities of network transport, such as packet packetization, and packet depacketization, while keeping massive network transport states, such as QP entries, on FPGA's on-board or/and on-chip memory. As such, servers enter NIC-centric coordination, rather than CPU-centric coordination that needs the CPU's network transport to consume each packet, as shown in Figure~\ref{fig_transport}b. The benefit of such an offloading design is that we can reduce the network transport time dramatically to \~2us because of the delicate customized hardware design, thus network transport and P4 switch have roughly the same latency level. 
Such an offloading design allows the host CPU to focus on handling complex application logic instead of wasting time and resources on complex network transport. 

%

%

%

\subsection{How to Complement an SSD?}
\label{subsec_ssd}
\vspace{-0.5ex}

We cannot always populate an entire huge dataset for a particular application, e.g., huge graph in the host memory due to its low capacity, e.g., hundreds of GB, so we have to explore how to leverage NVMe SSDs to increase memory capacity for data analytics on big data~\cite{ssd_vldb23, von_adms2022, much_icde2022, operation_sigmod2011, peak_icde2013, lruc_vldb2023, aceing_icde2023, hippogriffdb_vldb2016, merge_icde2017, bushstore_icde2024, latte_icde2020, rethinking_sigmod2020, scalestore_icde2022, tengine_icde2024, extended_icde2022, ldc_icde2019, dotori_vldb2023, better_vldbj2021, leaderkv_icde2024, boosting_icde2024, neos_icde2024, enhancing_icde2024, ginex_vldb2022, gts_sigmod2016, gids_vldb2024, hyperion_icde2025, helios_ppopp2025, tsplit_icde2022, lohan_icde2025, evaluating_icde2021}. 
Typically, NVMe SSD manipulation consists of a control plane and a data plane. The data plane is well optimized, because it employs a hardware DMA engine within SSD to transfer data between NVMe SSD and PCIe buffer, e.g., host CPU and GPU memory. The main challenge comes from heavy overhead of the control plane. In the following, we first discuss the issues when keeping the control plane on CPU, and second present the deep co-optimization of FPGA and SSD to efficiently implement the control plane on FPGA.

\begin{figure}[t]
    \subfloat[CPU manipulating SSDs]{
        \label{fig_fpgassd_cpu}
        \includegraphics[width=0.49\linewidth]{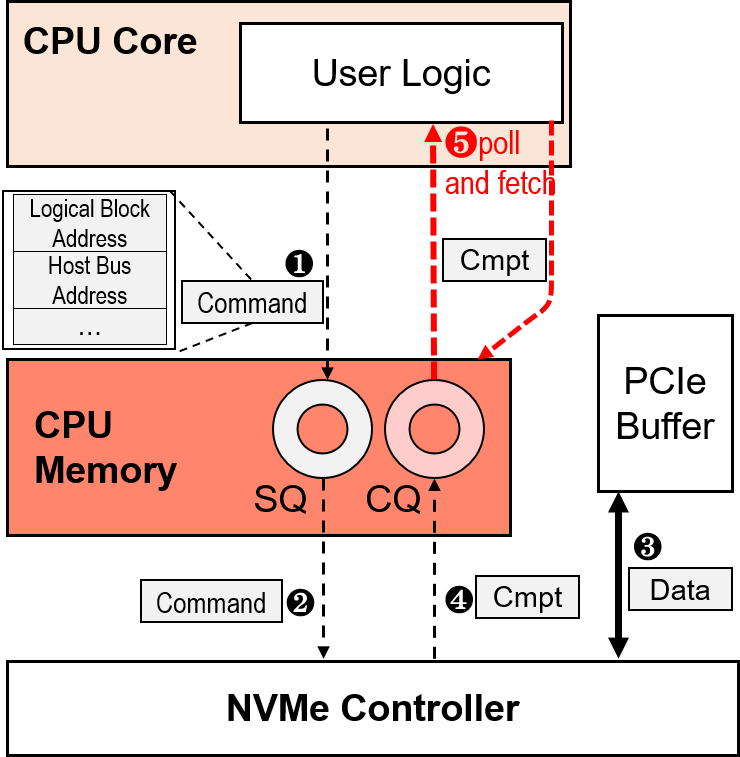}
    }
    \hfill    
    \subfloat[FPGA manipulating SSDs]{
        \label{fig_fpgassd_fpga}
        \includegraphics[width=0.44\linewidth]{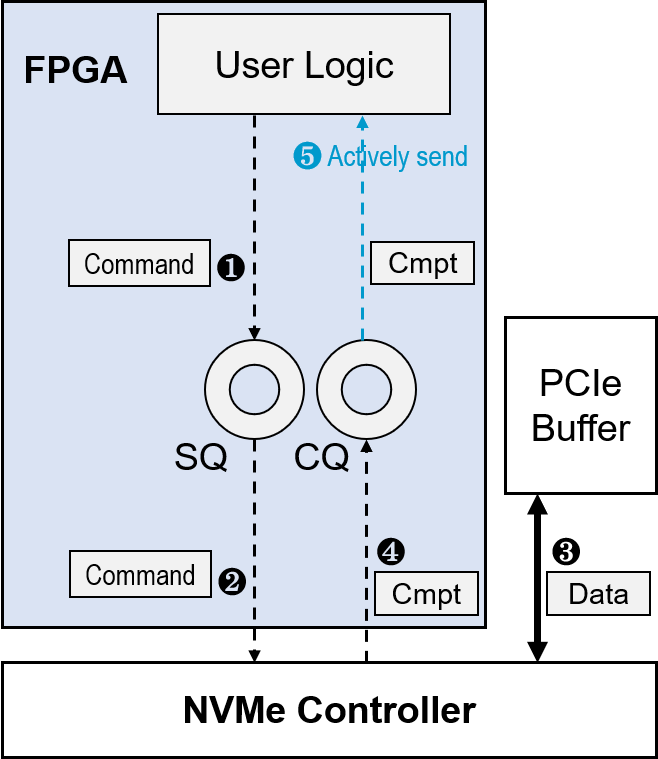}
    }
    \vspace{-1ex}
    \caption{CPU vs. FPGA: control plane of NVMe SSDs} 
    \vspace{-3ex}
    \label{fig_fpgassd} 
\end{figure} 

\subsubsection{Control Plane on CPU}
CPU interacts with NVMe SSDs mainly via two types of buffers, namely Submission Queues (SQs) and Completion Queues (CQs)~\cite{nvme_spec}. SQs and CQs are placed on CPU memory, and an NVMe controller stores a pointer to these buffers. Figure~\ref{fig_fpgassd_cpu} illustrates a typical paradigm of CPU manipulating NVMe SSDs, which consists of five steps: 
1)~CPU writes an NVMe command onto an SQ entry, which provides necessary information such as data direction, logical address of the data block on SSD, and PCIe bus address of host data buffer. Then, the CPU sends an SQ doorbell signal to the SSD. 
2)~After receiving the doorbell signal, the NVMe controller on the SSD fetches and processes the command from SQ. 
3)~The SSD performs data transfer via its DMA engine according to the command fields (data plane). 
4)~After the data transfer finishes, the SSD writes the completion information to a CQ entry. To be aware of the new completions, CPU has to poll each CQ during the process. 
5)~CPU reads and processes the completion from CQ, and sends a CQ doorbell signal to the SSD. 
We observe that polling the completion introduces high CPU overhead, thus reducing CPU efficiency. 

\subsubsection{Offloading Control Plane to FPGA} 
To relieve the issue regarding the CPU-based control plane, we offload the NVMe control plane to the FPGA. Figure~\ref{fig_fpgassd_fpga} illustrates an example of FPGA manipulating SSDs. 
1)~User logic writes an NVMe command onto an on-chip SQ entry. Then, the user logic sends an SQ doorbell signal to the SSD. 
2)~After receiving the doorbell signal, the NVMe controller on the SSD fetches and processes the command from SQ. 
3)~The SSD performs data transfer via its DMA engine according to the command fields (data plane). 
4)~After the data transfer finishes, the SSD writes the completion information to an on-chip CQ entry, of which the user logic actively captures its arrival. 

We identify two major differences in this design.  
First, we place SQs and CQs onto the FPGA's low-latency on-chip memory. It's natural for an FPGA to process multiple SQs and CQs simultaneously by instantiating SQ/CQ controlling units, each only requires a few hardware resources. In particular, FPGA logic can natively capture a CQ entry without involving to poll on expensive host memory from the CPU-based control plane, as shown in Figure~\ref{fig_fpgassd_fpga}. Besides, the NVMe controller exchanges commands and completions with the FPGA via peer-to-peer DMA, which enables FPGA to interact with SSDs directly without CPU's participation.  
Second, the data buffer is not limited to being on FPGA. It can also be on other PCIe devices such as GPU memory and CPU memory. The only difference introduced by different data buffer locations is the PCIe bus address field within an NVMe command. Unlike CPU, no modifications to file systems and kernel drivers are required. 

In summary, this decouples the control plane and data plane, thus enabling a more flexible system design while minimizing the overhead from the control plane.

\subsection{How to Complement a CPU?}
\label{subsec_cpu}
\vspace{-0.5ex}
Due to the relatively low performance of CPU, it is difficult and unrealistic to implement all functionalities on the CPU~\cite{dds_arxiv,cowbird_sigcomm23,smartds_isca23,ipipe_sigcomm19, enzian_asplos22}. Therefore, we need to discuss how to complement CPU with FPGA to process network traffic. In the following, we discuss three cases: control and data planes on CPU, control and data planes on the FPGA, and control plane on CPU and data plane on FPGA. 

\subsubsection{Control and Data Planes on CPU}
It is the most common design to use the CPU to implement both the control and data plane, because the short development cycles of CPU programming enable fast updates and fast evolution of high-level functionalities. Figure~\ref{fig:offloading}(a) shows the data flow when both the control plane and data plane are on CPU. 

However, the CPU's flexibility comes at the cost of low performance especially for compute-intensive functionalities, because both network message headers and network message payloads have to be processed by CPU, which would cause severe performance loss, especially for compute-intensive applications. Even worse, network speeds grow rapidly~\cite{nicmem_asplos22}, with 100 Gigabit Ethernet (GbE) network interface controllers (NICs) already widely available~\cite{broadcom_100G,mellanox_cx5}, 400 GbE arriving in 2021~\cite{broadcom_400G} and 800 GbE~\cite{eth_800G} expected in the near future. The growing network traffic would exacerbate the CPU's performance issue.

\subsubsection{Control and Data Planes on FPGA}
In order to address the above low performance issue, the naive solution is to offload all functionalities to FPGA, because the FPGA-only solution has high performance and lower active power due to its customized design~\cite{gimbal_sigcomm21}. Figure~\ref{fig:offloading}(b) shows the data flow when both the control plane and data plane are on the FPGA. As such, both network message headers and payloads have to be processed by FPGA logic. Because the payloads are usually required to be processed in a fixed manner, such a design can easily achieve high performance. 

However, simply offloading control plane to FPGA can bring new issues, because the control plane is usually very flexible and changeful~\cite{smartds_isca23, azure_accelnet,ali_solar}, so offloading the control plane to FPGA would inevitably sacrifice flexibility. It is unacceptable for control plane~\cite{smartds_isca23}, which usually needs to be flexible enough to provide high-level applications with the ability to evolve quickly. 


\subsubsection{Control Plane on CPU and Data Plane on FPGA}
\begin{figure}[t]
	\centering
	\includegraphics[width=\linewidth]{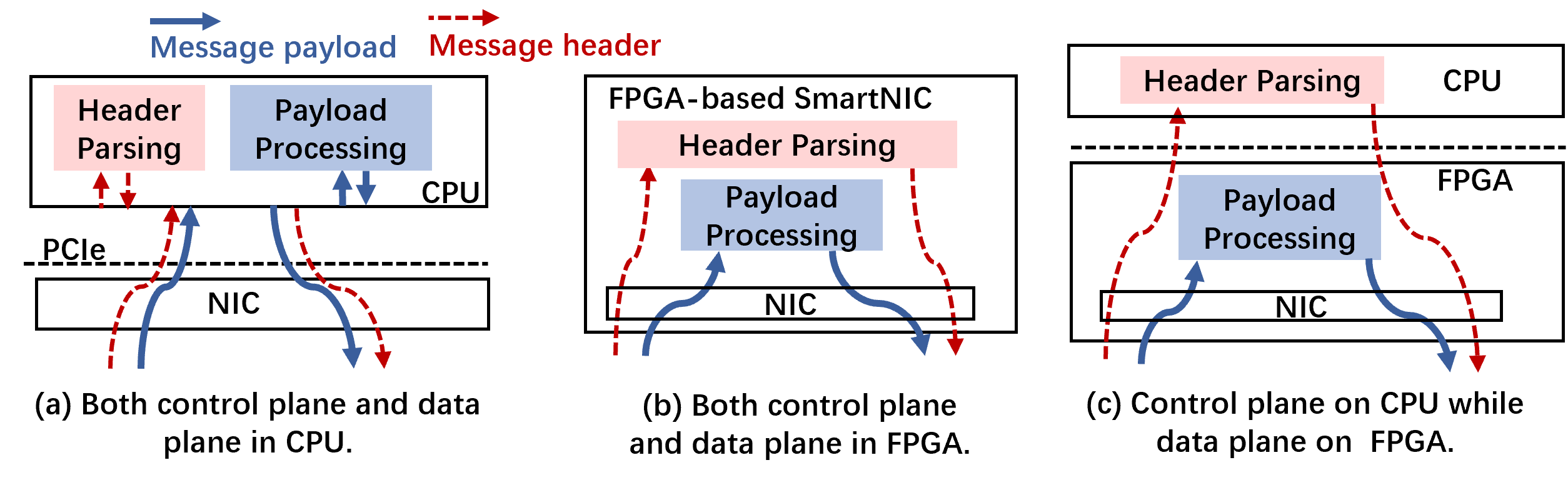}
	\vspace{-3ex}
	\caption{Comparison of CPU-FPGA co-processing methods.}
	\vspace{-3ex}
	\label{fig:offloading}
\end{figure}

We observe that the data plane usually requires fixed but heavy computations while the control plane requires flexible but lightweight computations, so we argue to keep the control plane on CPU while offloading the data plane to the FPGA, as shown in Figure~\ref{fig:offloading}(c). 
The FPGA transport layer would process each received network packet in FPGA memory. When all the packets of a network message arrive, FPGA will notify the CPU software of the coming of a new network message and forward the message header to the host CPU memory while the message payload remains in the FPGA memory. The message header size can be set in a per-flow manner and can vary according to the upper-layer applications.  
When sending a network message, the CPU software can notify the FPGA transport layer to reassemble its header from CPU memory and its payload from FPGA memory. 
As such, the control plane can enjoy the flexibility of CPU programming while the data plane can enjoy the high performance of hardware offloading.

%% file: content/3-Design.tex
\section{\sysname: Initial Design}

\begin{figure}[t]
	\centering
	\includegraphics[width=8.5cm]{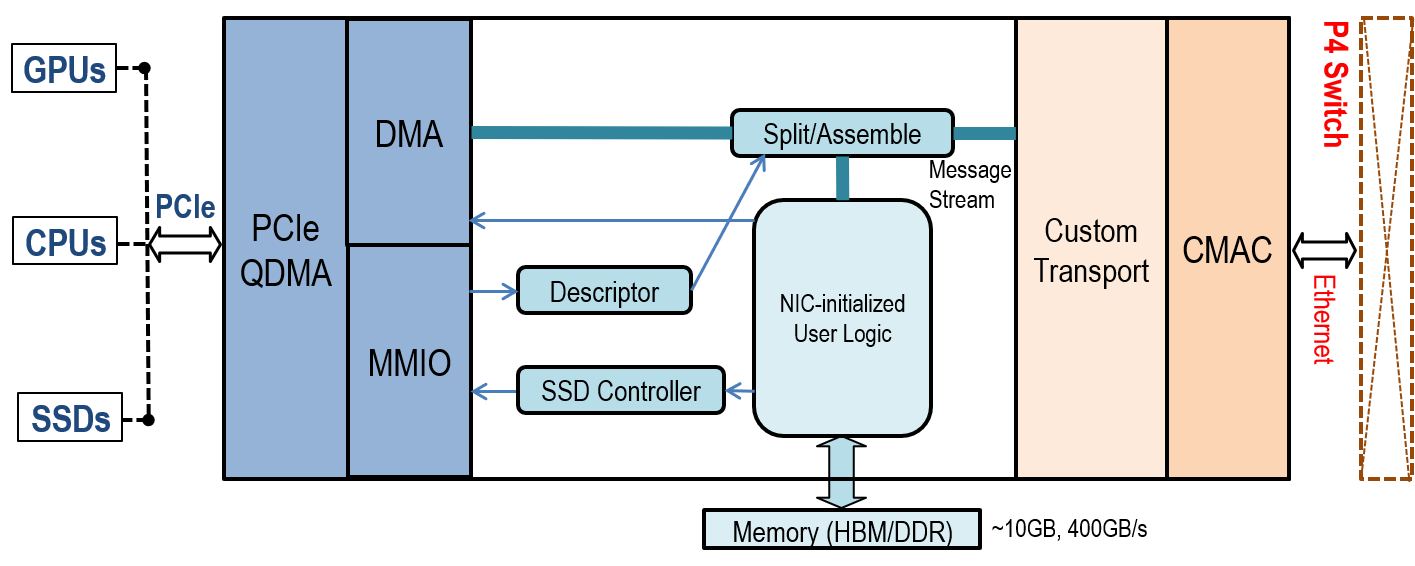}
	\vspace{-1ex}
	\caption{Initial Design of \sysname}
	\vspace{-3ex}
	\label{fig_fpga_design}
\end{figure}

In this section, we present the initial design of \sysname, which inspires the following hyper-heterogeneous research. 
\sysname{} argues for NIC-initiated design, rather than the traditional CPU-initiated design, for big data analytics that intends to leverage heterogeneous accelerators. To do so, \sysname{} mainly consists of three main components: PCIe, networking, and NIC-initiated user logic.


\noindent{\bf 1, PCIe Component. }
The PCIe IP core, in terms of QDMA, within an FPGA exposes DMA and memory-mapped IO (MMIO) for \sysname{} to exploit. The DMA engine allows \sysname{} to directly transfer a large amount of data to and from CPU/GPU memory without involving the host CPU. 
The MMIO exposes master and slave interfaces for us to explore. 
The MMIO master interface allows the host CPU/GPU to directly access the FPGA's internal registers or memory that are exposed as part of the system's memory space, such that CPU/GPU can directly manipulate \sysname. Specifically, we can instantiate one "descriptor" component via this interface to allow users to specify a user-defined descriptor to split/assemble each message on-demand via the ``split/assemble" component at runtime~\cite{smartds_isca23}. 
The MMIO slave interface allows \sysname{} to instantiate one "SSD Controller" component that needs only a small proportion of FPGA resources to directly manipulate tens of SSDs within a server, while saturating SSD bandwidth. 

\noindent{\bf 2, Networking Component. }
In order to provide flexibility, \sysname{} allows users to develop a custom network transport that works with the P4 switch to implement distributed primitives for data analytics applications. The custom network transport relies on the CMAC layer to provide a high-performance, low-latency Ethernet communication port that talks to external Ethernet devices, such as switch. 
The custom network transport provides \sysname{} a reliable message stream interface that directly connects to the ``split/assemble" component. This component relies on the corresponding descriptor from the ``descriptor" component to 1) split a message from the custom network transport to either the user logic or internal PCIe devices such as CPU or GPU, and 2) assemble a message from either the user logic or internal PCIe devices to the custom network transport.



\noindent{\bf 3, NIC-initiated User Logic. } 
The key part of \sysname{} is the NIC-initiated user logic component that initiates and then orchestrates the data flows between the network and the internal devices, such as GPU, CPU, and SSD. 
For example, this user logic, instead of the host CPU, can directly issue SSD operations on behalf of data analytics to fetch data from SSDs to the destination, once this module receives from the network a commend to access storage, so as to minimize long SSD access latency overhead. 
Moreover, this component features the on-board memory to host offloaded on-demand application states for better performance. 
Take LLM training and inference for example, we can offload the entire collective communication functionalities and states to \sysname, so as to fully overlap computation and communication, without wasting precious GPU resources~\cite{deepseekv3tr_arxiv24}.

%% file: content/4-experiment.tex
\section{Initial Experimental Evaluation}
\label{sec_experiment}
\vspace{-0.5ex}
\subsection{Experimental Setup}
\vspace{-0.5ex}
\noindent{\bf System Architecture. }We run the experiments on a cluster consisting of eight GPU+FPGA servers, connected by a Tofino P4 switch. Each server has two Intel Xeon Silver 4214 CPUs @2.20GHz, 128GB memory, a Xilinx UltraScale+ FPGA~\cite{xilinx_alveo}, several NVMe SSDs, and an Nvidia A100 GPU. 

\noindent{\bf Methodology. } 
Our key goal of the initial experiment is to demonstrate the key potential of \sysname, particularly how to use FPGA to complement each other heterogeneous device.  

\subsection{GPU + FPGA Co-design}

\begin{figure}[t]
    \subfloat[Control plane latency comparison. ]{
        \label{fig:contral_plane_latency}
        \includegraphics[width=0.45\linewidth]{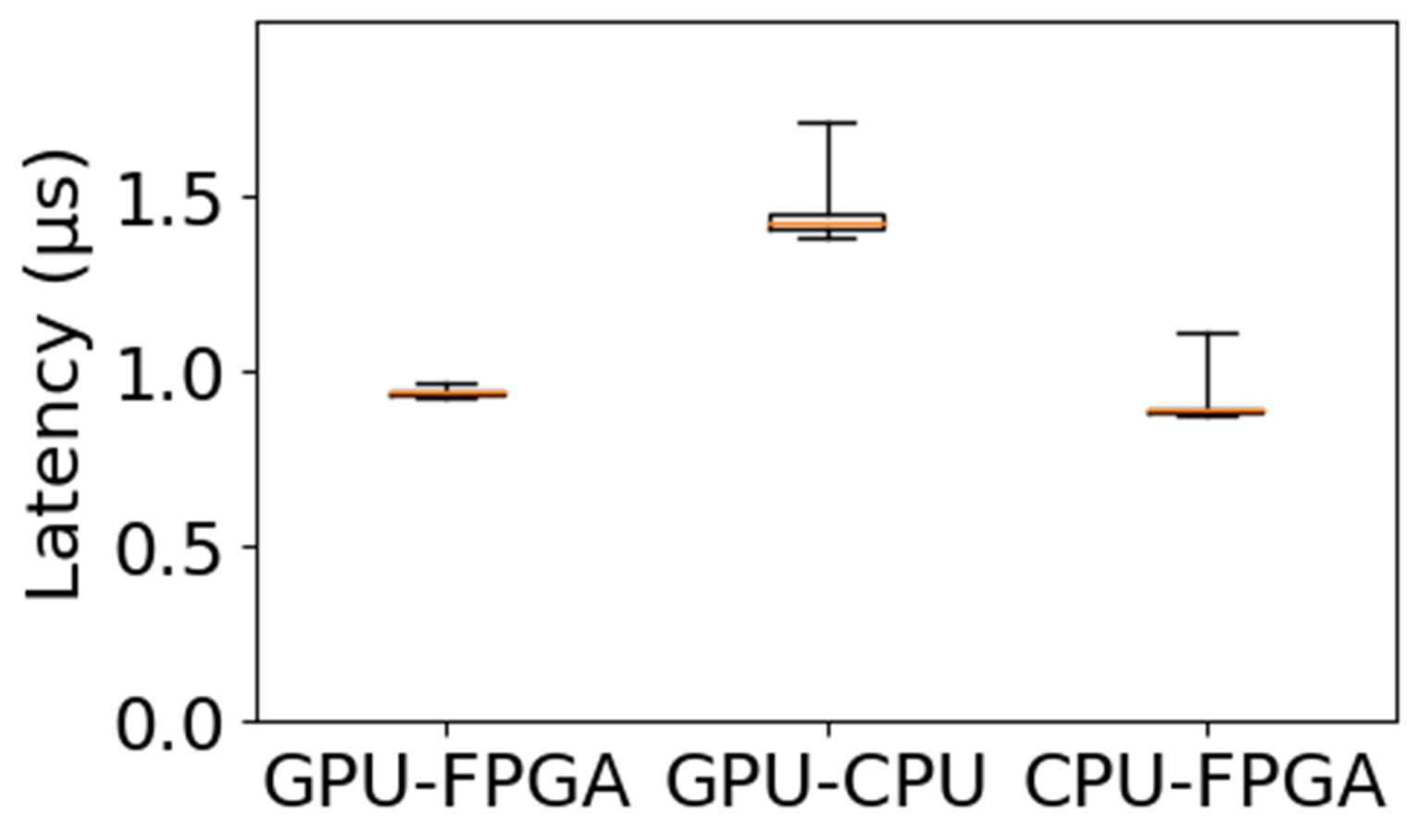}
    }
    \hfill    
    \subfloat[Cross-network inter-GPU latency]{
        \label{fig:contral_plane_time}
        \includegraphics[width=0.45\linewidth]{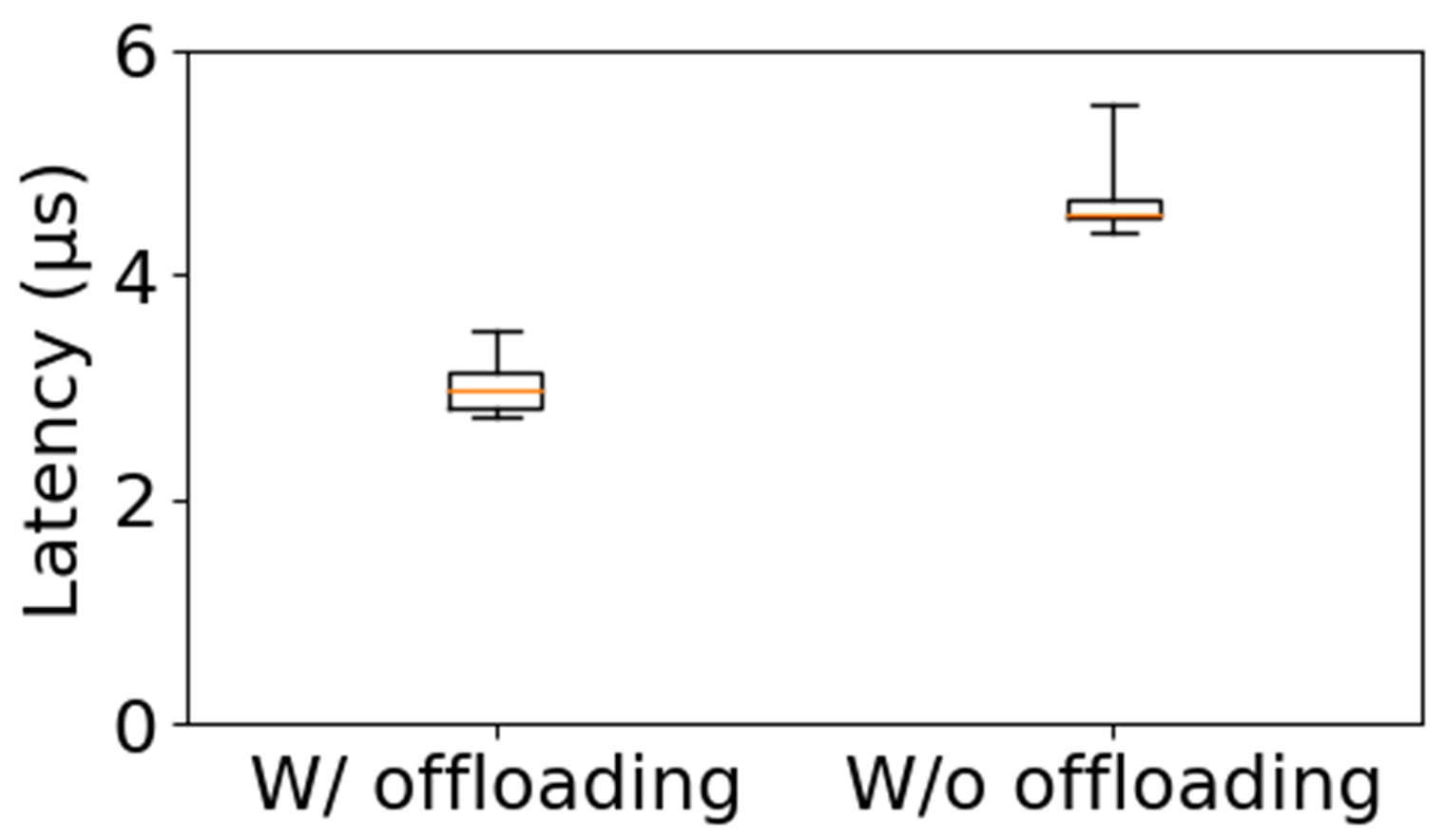}
    }
    \vspace{-2ex}
    \caption{With vs. without control plane offloading. ``X-Y'' refers to the device ``X'' accesses the device ``Y''.} 
    \vspace{-3ex}
    \label{fig:p4sgd} 
\end{figure} 

To validate the potential of \sysname-enhanced inter-GPU communication, we set up an experiment that allows one GPU to directly use load/store instructions within a CUDA kernel to communicate with a remote GPU through the direct model path: GPU-PCIe-FPGA-network-FPGA-PCIe-GPU, labeled as ``W/ offloading''. The baseline ``W/o offloading'' needs one GPU to inform its corresponding CPU, which relies on RDMA to talk to the other CPU. The other CPU sends the message to the remote GPU, so the data path is GPU-CPU(RDMA)-network-CPU(RDMA)-GPU.  
Figure~\ref{fig:contral_plane_latency} illustrates the read latency across different endpoints. ``X-Y'' indicates that device ``X'' reads from device ``Y''. We have two observations. First, GPU-FPGA has smaller latency fluctuations, compared with that of CPU-FPGA and CPU-GPU, demonstrating the advantage of \sysname{} in providing deterministic latency. Second, GPU-FPGA has significantly lower latency than the combined latency of GPU-CPU and CPU-FPGA, validating the efficiency of \sysname-enhanced inter-GPU communication.

Figure~\ref{fig:contral_plane_time} shows the cross-network inter-GPU latency comparison between with and without control plane offloading. We have two key observations. First, the case with control plane offloading achieves more stable latency compared to the case without offloading, highlighting \sysname{}'s advantage in achieving deterministic latency. Second, control plane offloading reduces the latency by approximately 50\%, due to the removal of kernel invocation overhead and context-switching bottlenecks, regardless of FPGA's low frequency. 


\subsection{P4 Switch + FPGA Co-design}
To validate the benefits of \sysname-enhanced P4 switch-centric design, we implement an efficient in-network aggregation primitive, where ``FPGA-Switch'' relies on the FPGA-based reliable network stack to send partial activations to the P4 switch that broadcasts the aggregated results back to all FPGAs. The CPU-Switch-based baseline~\cite{switchml} (labeled ``CPU-Switch'') relies on the CPU network stack to trigger RICs to send partial activations from the CPU memory to the P4 switch, which broadcasts the aggregated results back to all CPUs via NICs. 



\begin{figure}[t]
    \begin{minipage}{0.446\linewidth}
        \vspace{-1ex}
        \begin{center}
            \includegraphics[width=\linewidth]{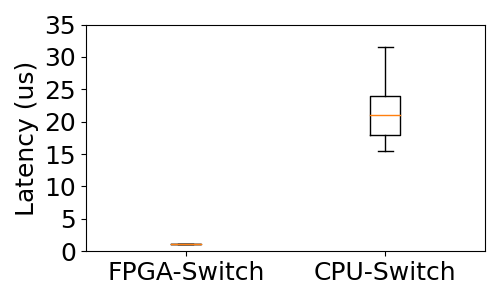}
        \end{center}
        \vspace{-1ex}
        \caption{\label{fig:p4sgd_latency} FPGA-Switch co-design} 
        \vspace{-3ex}
    \end{minipage} 
    \hfill
    \begin{minipage}{0.494\linewidth}
        \vspace{-1ex}
        \begin{center}
            \includegraphics[width=\linewidth]{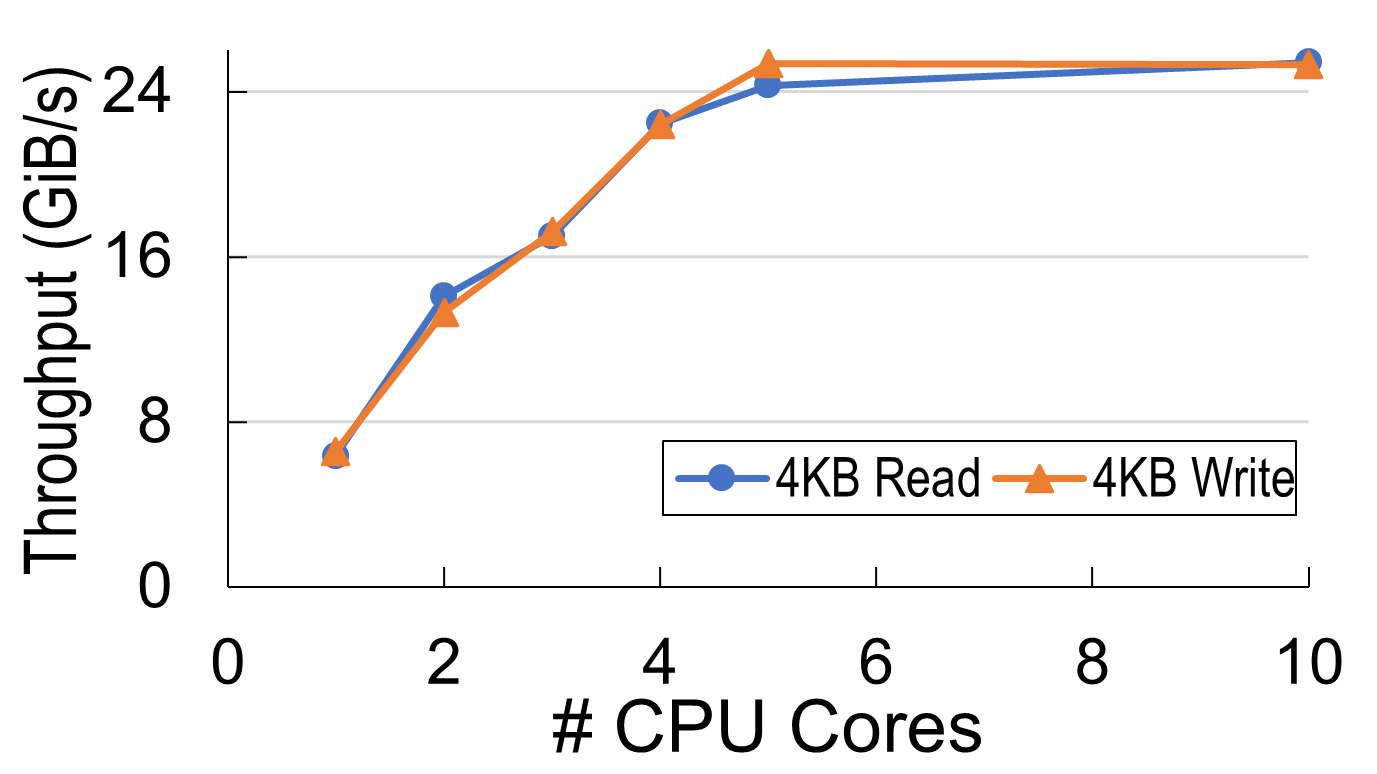}
        \end{center}
        \vspace{-1ex}
        \caption{\label{fig_fpgassd_cpucore} Throughput of CPU-based SSD control plane}
        \vspace{-3ex} 
    \end{minipage} 
\end{figure}

Figure~\ref{fig:p4sgd_latency} illustrates the latency comparison. We observe ``FPGA-Switch'' is able to reach an average latency of 1.2$\mu$s, which is an order of magnitude smaller than that of ``CPU-Switch". It's mainly because \sysname{} achieves deterministic low transport latency due to its customized hardware design in near-network position, while ``CPU-Switch" suffers from two severe overheads: 1) long cross-PCIe communication, and 2) expensive CPU software cost, as shown in Figure~\ref{fig_transport}. We envision that offloading the entire network transport to FPGA can significantly accelerate the applications that rely on low coordination cost. 
%

\begin{table}[t]\small
    \centering
    \caption{Resource usage of FPGA-based SSD control logic}
    \vspace{-2ex}
    \label{table_fpgassd_resource}
    \begin{tabular}{cccc}
        \specialrule{.1em}{.05em}{.05em} 
        \textbf{LUT} & \textbf{FF} & \textbf{BRAM} & \textbf{URAM} \\ \specialrule{.1em}{.05em}{.05em} 
        \makecell{45K \\ (5.2\%)} & \makecell{109K \\ (6.3\%)} & \makecell{164 \\ (12.2\%)} & \makecell{2 \\ (0.3\%)} \\ 
        \specialrule{.1em}{.05em}{.05em} 
    \end{tabular}
    \vspace{-4ex}
\end{table}

\subsection{SSD + FPGA Co-design}
To show the benefits of SSD+FPGA co-design, we show the hardware resource usage of CPU- and FPGA-based SSD control plane design. In the following, we first evaluate the CPU core usage of CPU-based design, then evaluate the hardware resource usage of FPGA-based design. 

To evaluate the CPU core usage of the CPU-based SSD control plane, we measure the I/O throughput of our design under different CPU core numbers. We run the experiments on an Intel Xeon Gold 5320 CPU @ 2.20 GHz and 10$\times$ D7-P5510~\cite{p5510} SSDs. The workload is 4 KB random read/write. The CPU uses SPDK~\cite{spdk}, a user-space storage I/O library to manipulate SSD access. Each CPU core directly generates and handles the I/O commands without any other workloads such as indexing.\footnote{Current benchmarking does not involve any workload. If any workload is added, more CPU cores are needed due to the introduced dependency that affects concurrency. } 
Figure~\ref{fig_fpgassd_cpucore} shows the throughput under different CPU core numbers. We observe that it requires 5 CPU cores to saturate the SSDs' I/O bandwidth for both read and write workloads. In contrast, the FPGA-based SSD control plane eliminates the CPU's participation. 

To evaluate the hardware resource usage of the FPGA-based SSD control plane, we implement the SSD control logic on an Alveo U50 FPGA~\cite{u50}, Table~\ref{table_fpgassd_resource} lists the resource usage when manipulating 10 SSDs. The SSD control module only costs 12.2\% of BRAMs and less than 10\% of other types of resources of a U50 board. We conclude that our FPGA-based SSD control plane is lightweight so that an FPGA can further integrate functions such as networking, compression/decompression, and encryption/decryption to further cooperate with SSD I/O. 

\subsection{CPU + FPGA Co-design}
To show the benefits of CPU-FPGA co-optimization, we run a realistic cloud block storage middle-tier application~\cite{smartds_isca23}. The application 1) receives the storage write requests from the computing servers; 2) compresses the payload in the request dynamically; and 3) sends the result into three disk servers which would store the result into disks. The baseline ``CPU-only'' runs the entire application on the CPU. ``CPU-FPGA'' implements the most compute-intensive logic, i.e., compression, on the FPGA while the remaining light logic runs in the host CPU. 

\begin{figure}[t]
    \subfloat[Throughput]{
        \label{fig_smartds_throughput}
        \includegraphics[width=0.47\linewidth]{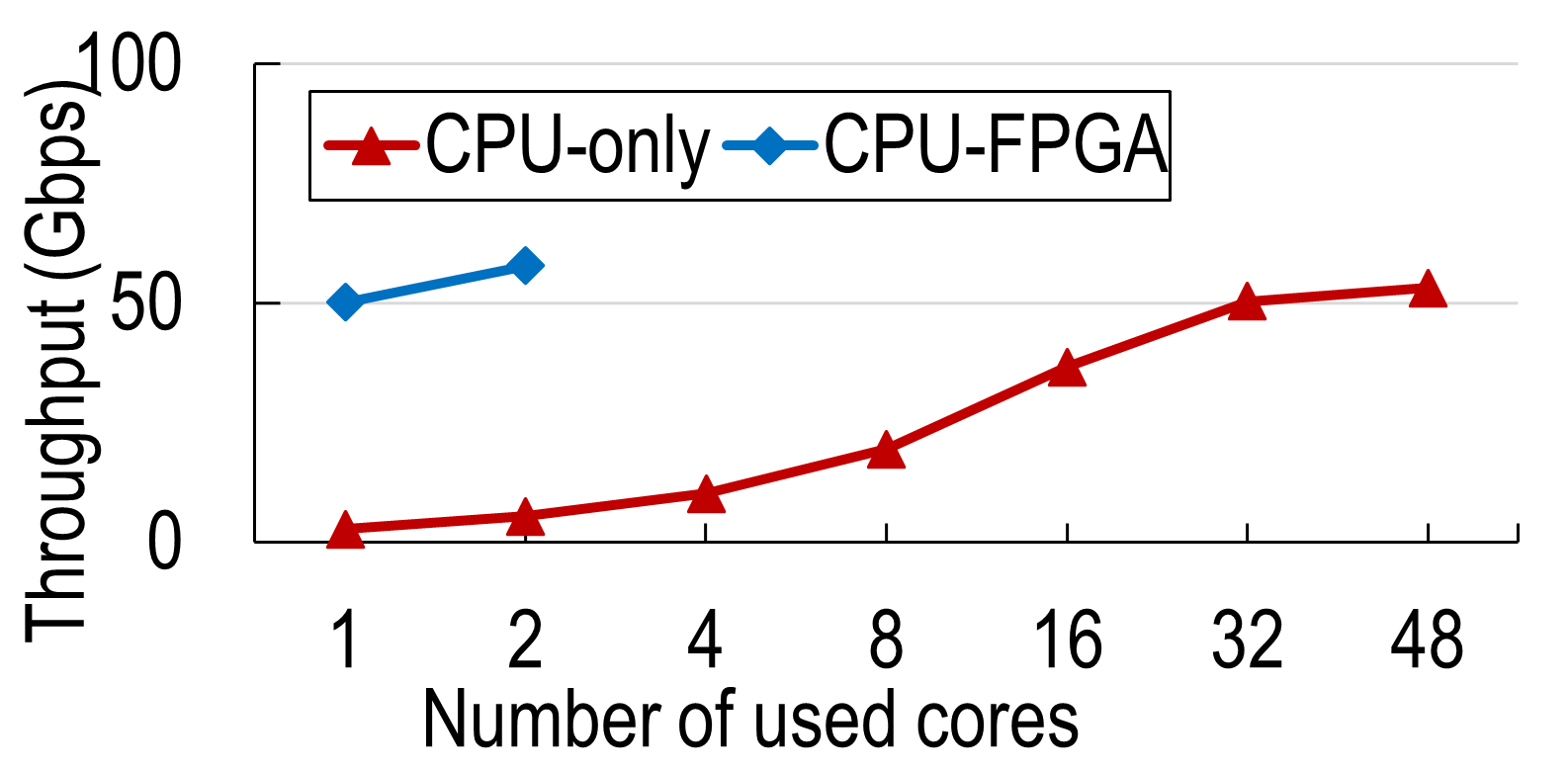}
    }
    \hfill    
    \subfloat[Average Latency]{
        \label{fig_smartds_latency}
        \includegraphics[width=0.47\linewidth]{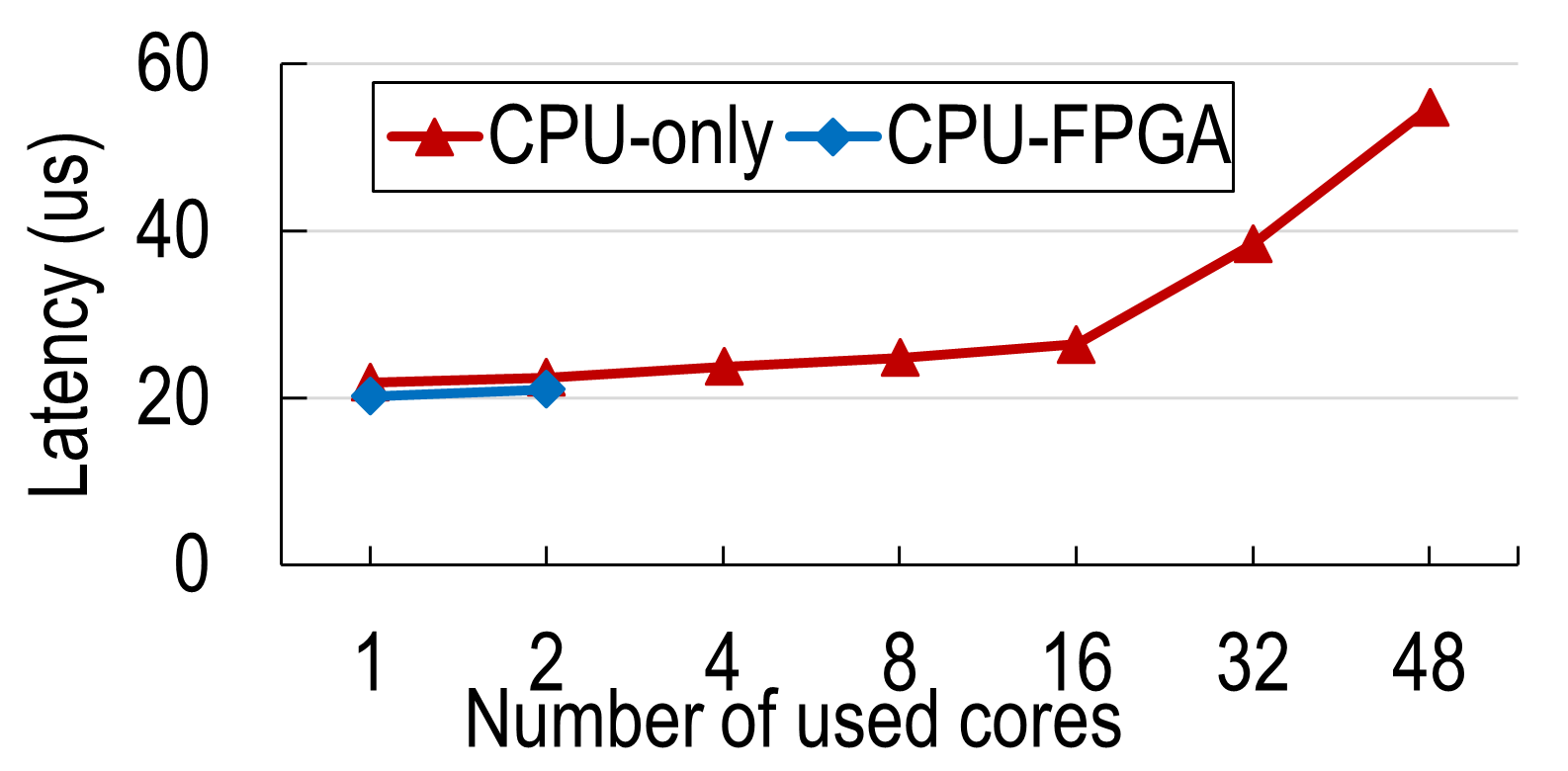}
    }
    \vspace{-1ex}
    \caption{CPU-only vs. CPU-FPGA co-processing} 
    \vspace{-3ex}
    \label{fig_smartds} 
\end{figure}

Figure~\ref{fig_smartds_throughput} shows the achievable throughput when varying the number of used CPU cores. We observe that ``CPU-FPGA'' only requires two CPU cores to achieve the maximum processing throughput while ``CPU-only'' requires all 48 cores, because CPU-based compression is a compute-bound task, and a single core can only achieve 1.6 Gbps LZ4 compression throughput. In contrast, hardwired compression is very easy to achieve high throughput in FPGAs. 
Figure~\ref{fig_smartds_latency} shows the average latency. We observe that ``CPU-FPGA'' always keeps a relatively low latency, while the latency ``CPU-only'' greatly increases when the number of used cores increases, because the majority of a message, i.e., payload, is processed at the FPGA, rather than on the CPU. 
In summary, \sysname{} enables us to enjoy flexible CPU programming on the control plane while achieving high-throughput data plane processing on the FPGA.

%% file: content/5-conclusion.tex
\section{Conclusion}
\label{sec_conclusion}
	\vspace{-1ex}
In this paper, we present \sysname, an FPGA-centric, hyper- heterogeneous computing platform for big data analytics. The key idea of \sysname{} is to position FPGAs as a hub that complements other heterogeneous devices, based to its highly reconfigurable nature and rich IO interfaces such as PCIe, networking, and on-board memory. As such, \sysname{} enables fine-grained interplay between network, storage, and computing power, and enables architectural flexibility to allow a huge design space exploration on our hyper-heterogeneous computing platform, with the key goal of ``1+1>2".  